\documentclass{article}
\usepackage{multicol}
\usepackage[dvips]{graphicx}
\usepackage{amsmath}
\usepackage{epsfig}
\newcommand{\bfr}{\begin{flushright}}
\newcommand{\efr}{\end{flushright}}
 
\begin{document}
\title{A New $2+1$ Dimensional Einstein Gravity Solution Coupled to
Born-Infeld Electromagnetic Theory without Cosmological Constant 
}
\author{Masashi Kuniyasu\footnote{s007vc@yamaguchi-u.ac.jp}
\\
{\small
Graduate School of Science and Technology, Yamaguchi University,}\\
{\small Yamaguchi-shi, Yamaguchi 753--8512, Japan}
}
\date{\today
}
\maketitle
\begin{abstract}
A new solution of the Einstein-Born-Infeld theory in $2+1$
space-time is derived. A new solution has no horizon there are
two singularity. This space-time has two singular points, however,
one of the point at the origin is not in the physical region.
We also investigate the energycondition. Then, weak energy condition
is satisfied. However, causal energy condition is violated.
\\
~
\\
\end{abstract}
\begin{multicols}{2}

The theory of gravity in $2+1$ dimensional space-time was first
investigated by Deser et al~\cite{SR}. Anyone could not find
black hole solution in $2+1$ space-time around 80's. However,
Ba\~nados et al~\cite{BTZ} derived black hole solution in the $2+1$
Anti-de-Sitter space time. After their research, the solution in $2+1$
space-time with matter field was investigated. In this paper, we attention
non linear electrodynamic (NED) as the matter field.

Einstein's gravity solutions in $2+1$ space-time that couple to
NED was first investigated by Caldro et al~\cite{CA},\cite{CG}. They treated
circular symmetric electric field. Recently, Mazaharimousavi found new
solutions with special vector potential that is considered in
our paper. There is consistent solutions with Maxwell
electrodynamic field~\cite{MMH} and $\sqrt{|F|}$ type NED without cosmological
constant~\cite{MH}. Then, in this paper, we try to find a new $2+1$ gravity
solution coupled to Born-Infeld type NED\cite{BI} without cosmological constant.

Let us take the $2+1$ dimensional action which is coupled to
electrodynamic field without cosmological constant
\begin{equation}
I=\frac{1}{2}\int d^3x\sqrt{-g}(R-\alpha L(F))\,.
\label{eq1}
\end{equation}
Where $R$ is the Ricci scalar, $\alpha$ is a coupling constant
and $F$ is the Maxwell invariant which is defined by $F=F_{\mu\nu}F^{\mu\nu}$
with $F_{\mu\nu}=\partial_{\mu}A_{\nu}-\partial_{\nu}A_{\mu}$. The explicit
form of the Laglangian is given by
\begin{equation}
L(F)=4b^2\left(1-\sqrt{1+\frac{F}{2b^2}}\right)\,.
\label{eq2}
\end{equation}
Where $b$ is a constant. We get the following Einstein equation from the
variation of the action (\ref{eq1}) which is respect to $g_{\mu\nu}$
\begin{equation}
G_\nu^\mu=T_\nu^\mu\,.
\label{eq3}
\end{equation}
Where $T_{\mu\nu}$ is the stress tensor of the electromagnetic field
given by
\begin{equation}
T_\nu^\mu=\alpha\left(-2F^{\mu\lambda}F_{\nu\lambda}L_{,F}+
\delta_{\nu}^{\mu}\frac{L}{2}\right)\,,
\label{eq4}
\end{equation}
in which $L_{,F}$ means differential of $F$ and the explicit form is
\begin{equation}
L_{,F}=\frac{-1}{\sqrt{1+F/2b^2}}\,.
\label{eq5}
\end{equation}

Next, let us set the electric field ansatz as
\begin{equation}
F_{t\theta}=E_0\,.
\label{eq6}
\end{equation}
Where $E_0$ is a constant. Electric potential such as
\begin{equation}
{\bf A}=E_0\left( -c\theta ,0,at\right)\,,
\label{eq7}
\end{equation}
generate the electric field (\ref{eq6}) (two constant $a$ and $c$
satisfy the relation $a+c=1$). Above electric field
ansatz was investigated by Mazharimousavi at el~\cite{MMH},\cite{MH}.
They considered $F^k$ type electromagnetic field of $k=1/2,1$.
Then, we set metric ansatz as follows:
\begin{equation}
ds^2=-\left(\frac{r}{r_0}\right) ^2dt^2+B(r)dr^2+r^2d\theta ^2\,.
\label{eq8}
\end{equation}
In which $r_0$ is a constant, and $B(r)$ is some arbitrarily function.
Such as metric is derived by $F^{1/2}$ type electromagnetic field
(radial scalar field without a cosmological constant is also consistent
with above metric ansatz~\cite{SS}).
So, we try (\ref{eq8}) type ansatz. This metric generate the
following Einstein tensor
\begin{equation}
G_t^t=G_{\theta}^{\theta}=-\frac{1}{2rB^2}\frac{dB}{dr}\,,
\label{eq9}
\end{equation}
\begin{equation}
G_r^r=\frac{1}{r^2B}\,.
\label{eq10}
\end{equation}
Addition, Maxwell invariant $F_{t\theta}$ is given by
\begin{equation}
F=2F_{t\theta}F^{t\theta}=-2\frac{r_0^2E_0^2}{r^4}\,.
\end{equation}
Non zero component of the stress tensor becomes as follows
\begin{equation}
T_t^t=T_{\theta}^{\theta}=-2\alpha b^2\frac{\left(1-\sqrt{1-\frac{r_0^2
E_0^2}{r^4b^2}}\right)}{\sqrt{{1-\frac{r_0^2E_0^2}{r^4b^2}}}}\,,
\label{eq11}
\end{equation}
\begin{equation}
T_r^r=2\alpha b^2\left(1-\sqrt{1-\frac{r_0^2E_0^2}{r^4b^2}}\right)\,.
\end{equation}

From radial component of the Einstein equation (\ref{eq3}), we can
determinate function $B(r)$
\begin{equation}
B(r)=\frac{1}{2\alpha b^2r^2\left(1-\sqrt{1-\frac{r_0^2E_0^2}
{r^4b^2}}\right)}\,.
\label{eq13}
\end{equation}
This function is consistent with temporal and angle part of the
Einstein equation because the differential of the function $B(r)$ is
\begin{equation}
\frac{dB}{dr}=4\alpha b^2rB^2\frac{\left(1-\sqrt{1-\frac{r_0^2
E_0^2}{r^4b^2}}\right)}{\sqrt{{1-\frac{r_0^2E_0^2}{r^4b^2}}}}\,.
\label{eq14}
\end{equation}
That is why we conclude the new $2+1$ Einstein-Born-Infeld
solution without the cosmological constant is
\begin{equation}
\begin{split}
ds^2=&-\left(\frac{r}{r_0}\right) ^2dt^2 \\
     &+\frac{dr^2}{2\alpha b^2\left(r^2-\sqrt{r^4-\frac{r_0^2E_0^2}
	  {b^2}}\right)}+r^2d\theta ^2\,.
\label{eq15}
\end{split}
\end{equation}
We notice that $r$ must be lager than $\sqrt{r_0E_0/b}$. Addition,
$\alpha$ must be positive to get the $2+1$ space-time.

Next, to analyze the singular point of space-time (\ref{eq15}), we calculate
the invariant ${\cal R} = R_{\mu \nu \lambda \sigma}
R^{\mu \nu \lambda \sigma}$
\begin{equation}
\begin{split}
{\cal R}=&2\left( \frac{1}{rB^2}\frac{dB}{dr}\right) ^2+\frac{4}{r^4B^2} \\
		=&8(G_t^t)^2+4(G_r^r)^2\,.
\label{eq16}
\end{split}
\end{equation}
Here, let us use the Einstein equation (\ref{eq3}) and we get
\begin{equation}
\begin{split}
{\cal R}=&16\alpha^2b^4\left(r^2-\sqrt{r^4-\frac{r_0^2E_0^2}{b^2}}\right)^2\\
         &\times\frac{1}{r^4}\frac{3r^4-r_0^2E_0^2/b^2}{r^4-r_0^2E_0^2/b^2} \,.
\label{eq17}
\end{split}
\end{equation}
From the invariant ${\cal R}$, this space time has two singular points
at $r=0$ and $r=\sqrt{r_0E_0/b}$. However, physical region of space-time
(\ref{eq15}) is $r>\sqrt{r_0E_0/b}$, $r=0$ singular point is not naked.
On the other hand, singular $r=\sqrt{r_0E_0/b}$ is naked. That is why
we conclude our new solution (\ref{eq15}) has one naked singular point.

Finally, let us consider the energy condition. We define following
quantities;
\begin{equation}
\rho=-T_t^t\,,
\end{equation}
\begin{equation}
p=T_r^r\,,
\end{equation}
\begin{equation}
q=T_{\theta}^{\theta}=-\rho\,.
\end{equation}
Weak energy conditions (WECs) means, i)~$\rho\ge0$,~ii)$\rho+p\ge
0$~and~iii)$\rho+q\ge0$. Then,
\begin{equation}
\rho+p=\frac{r_0^2}{r^2}\frac{2\alpha E_0^2}
{\sqrt{r^4-r_0^2E_0^2/b^2}}\,,
\end{equation}
\begin{equation}
\rho+q=0\,.
\end{equation}
That is why WECs are all satisfied if the coupling $\alpha$ is
positive. It is consistent with the spatial part of the metric is
larger than zero. This condition is also derived by Maxwell type
theory\cite{MMH}. On the other hand,
\begin{equation}
\begin{split}
&p_e=\frac{p+q}{2} \\
&=-\alpha b^2\left[\left(1-\frac{r_0^2E_0^2}{r^4b^2}
\right)^{-1/4}-\left(1-\frac{r_0^2E_0^2}{r^4b^2}\right)^{1/4}\right]^2\,.
\end{split}
\end{equation}
It means the causal energy condition $0\le p_e \le1$ will violate.
Moreover, if we take $E_0/b\sim0$, $p_e$ reaches to zero and the causal
energy condition will be satisfied. This means Maxwell limit of the 
Born-Infeld theory which is consistent with condition of the Einstein
Maxwell solution\cite{MMH}.

In this article, we derived a new solution of the
Einstein-Born-Infeld theory without cosmological constant.
A new solution (\ref{eq15}) has no horizon and singular point lies
at $r=0$ and $r=\sqrt{r_0E_0/b}$. However, physical region of the 
space-time (\ref{eq15})is larger than $\sqrt{r_0E_0/b}$.
So, we conclude there is one naked singularity in the new 
space-time (\ref{eq15}). We also investigated energy condition,
and weak energy conditions was satisfied. However, causal energy
condition was violated. Moreover, in the Maxwell limit,
causal energy condition will recovered.

\section*{Acknowledgments}

I would like to thank Doctor K.~Shiraishi for helpful discussions.
I also thank him for reading the manuscript.

\end{multicols}

\newpage


\bibliographystyle{apsrev4-1}



\end{document}